\documentclass{kluwer}
\usepackage{epsfig}

\def\ga{\mbox{ \raisebox{-.5ex}{$\stackrel{\textstyle >}{\sim}$} }}

\begin{document}                                                                                   
\begin{article}
\begin{opening}

\title{Temporal Correlation of Hard X-rays and Meter/Decimeter Radio Structures in Solar Flares} 

\author{Kaspar \surname{Arzner}$^1$ and Arnold O. \surname{Benz}$^2$}  
\runningauthor{Kaspar Arzner and Arnold Benz}
\runningtitle{Correlation of HXR and m/dm Radio Structures}

\institute{$^1$Paul Scherrer Institut, CH-5232 Villigen PSI, Switzerland \\
$^2$Institute of Astronomy, ETH Z\"urich, CH-8092 Zurich, Switzerland}

\date{\today}

\begin{abstract}
We investigate the relative timing between hard X-ray (HXR) peaks and structures in metric and decimetric radio emissions of solar 
flares using data from the RHESSI and Phoenix-2 instruments. The radio events under consideration 
are predominantly classified as type III bursts, decimetric pulsations and patches. The RHESSI data
are demodulated using special techniques appropriate for a Phoenix-2 temporal resolution of 0.1s. The absolute timing accuracy of 
the two instruments is found to be about 170 ms, and much better on the average. It is found that type
III radio groups often coincide with enhanced HXR emission, but only a relatively small fraction 
($\sim$ 20\%) 
of the groups show close correlation on time scales $<$ 1s. If structures correlate, the HXRs precede the type III emissions in a majority of cases, 
and by 0.69$\pm$0.19 s on the average. Reversed drift type III bursts are also delayed, but high-frequency and harmonic emission is 
retarded less. The decimetric pulsations and patches (DCIM) have a larger scatter of delays, but do not have a statistically 
significant sign or an average different from zero. The time delay does not show a center-to-limb variation excluding simple 
propagation effects. The delay by scattering near the source region is suggested to be the most efficient process on the average 
for delaying type III radio emission. 
\end{abstract}


\end{opening}

\section{Introduction}
 
A large fraction of the energy released in solar flares first appears in accelerated particles. Energetic electrons 
generate both impulsive radio emissions and hard X-rays (HXR), which are often closely associated. The HXRs, emitted
by electron-ion bremsstrahlung, were occasionally found to have temporal fine structures down to several 10 ms \cite{dennis85}, 
but more often (in some 10\% of all $\ge M$ class events) to the order of several 100 ms \cite{kiplinger84}. A very tight 
correlation is regularly observed between the radio emission produced incoherently by the synchrotron mechanism in centimeter 
wavelengths and bremsstrahlung X-rays as high-energy electrons are involved in both. Less correlation or even no association 
has been reported for the coherent radio emissions by electron beams, trapped electrons and from less-known mechanisms 
involving probably also non-thermal electrons \cite{benz05}. This is likely to be caused by the limited sensitivity of
present HXR observations.

Among the different types of coherent radio emission in the meter and decimeter ranges, type III bursts, narrowband spikes and 
pulsations often concur with HXRs. As both HXR and radio emissions are emitted presumably by nonthermal electrons,
the question arises whether both originate from the same electron population. More than in a general association, 
e.g. with type III groups, such an identity would be manifest in correlations of individual structures. 

Indeed, earlier observations revealed occasional correlations between individual type III bursts and HXR pulses
(e.g., at meter waves \opencite{Kane82}, and at decimeter waves \opencite{benz83}, \opencite{aschwanden95a}) at a timing accuracy of a few 
0.1 seconds.  Type III bursts are caused by electron beams exciting Langmuir waves in the coronal plasma. \inlinecite{Dennis84} found 
coincidences of reversed-slope type III bursts drifting downward in the corona and HXR peaks. 
A linear relation between the rate of type III bursts per second and the HXR emission has been reported for a case including a 
rich group of radio bursts (\opencite{aschwanden95b}). The frequency range of the above comparisons was limited to less than 1 GHz, 
and photon energies to more than 25 keV. None of the reported coincidences were without time shifts of the order of a few 0.1 s. 
As there are many reasons for delays of one or the other emission, including clock errors, the small difference did not cause 
much concern.


Good correlation between the integrated flux of narrowband spikes in decimeter radio waves and the HXR flux has been reported 
by \inlinecite{BenzKane86} and \inlinecite{Guedel91}. However, \inlinecite{Aschwanden92} noted that the integrated spike radiation is 
delayed by 1 - 2 seconds. Decimetric pulsations are more frequent than narrowband spikes. Some correlation with hard X-rays 
has been noticed by \inlinecite{benz83} and \inlinecite{Aschwanden85}. \inlinecite{Kliem00} and \inlinecite{Farnik03} reported a 
detailed anti-correlation or a delay of the fine structures in pulsating radio emission. Narrowband spikes and pulsations
are generally believed to originate from velocity-space instabilities, such as caused by a loss-cone distribution.
However, there is no confirmed theory at present.

The present investigation searches systematically for possible radio-HXR correlations on sub-second time scales using data from 
the Phoe\-nix-2 (\opencite{benz91}, \opencite{messmer99}) and RHESSI \cite{lin02} experiments. The availability of this new radio 
data set allows us to extend existing studies towards higher frequencies that include coherent emissions of denser sources and 
that are possibly more relevant for flare physics. The greatly enhanced spectral resolution of RHESSI allows including lower 
energy photons, selecting the most relevant energy range, and separating it from thermal emissions. As RHESSI is spin 
modulated, time resolution below the spin period requires careful demodulation, described in the second chapter. The demodulation 
introduces a relative timing uncertainty, which is however offset by the order of magnitude gain in absolute accuracy compared to previous 
observations.

\section{Data selection, reduction, and preprocessing}
 
From the period of February 2002 to March 2003, a set of events was selected according to the 
following criteria (in this order): (i) Presence of distinct radio structures on time scales below 1 s. 
(ii) A minimum simultaneously observed HXR flux of 50 ct/s/subcollimator above 6 keV. 
(iii) Absence of RHESSI attenuator movements during the investigated 
contiguous time range (typically, about one minute), and (iv) the availability of an approximate source position, 
which is needed for demodulation (see below). Out of the 40 initial candidates, only 22 revealed HXR-radio correlation
on inspection by eye, and were further investigated. The investigated events are compiled in Table \ref{eventlist_tab}.

\begin{table}[ht]
\begin{tabular}{cccccc}
Time          & Frequency       & Energy       & Delay  & Radio  \\
 
[UTC]		   & [MHz]	 & [keV] 	 & [s] 	  & Type  \\\hline
14-Feb-02 11:05:16 &  230 -  659 &     9 -    57 &   -0.84  &  III \\
14-Feb-02 11:06:01 &  452 - 1240 &     9 -    57 &    0.35  &  III \\
17-Mar-02 10:16:19 &  983 - 3820 &    18 -    57 &   -0.87  &  DCIM \\
15-Apr-02 08:51:35 &  381 -  667 &    10 -    70 &   -0.59  &  III \\
15-Apr-02 08:52:42 &  417 -  587 &    11 -    84 &   -0.43  &  III \\
20-May-02 10:50:38 &  569 - 1360 &    18 -    81 &    0.89  &  III \\
20-May-02 10:52:36 &  399 - 1600 &    31 -   194 &   -1.22  &  III U,RS \\
03-Jun-02 14:43:24 & 2500 - 3640 &    18 -    74 &    0.01  &  DCIM \\
03-Jun-02 17:13:32 &  524 - 3610 &    15 -    70 &   -1.97  &  III RS \\
01-Aug-02 07:42:30 &  246 -  515 &    20 -    72 &   -1.58  &  III \\
01-Aug-02 07:43:25 &  426 -  623 &    20 -    72 &   -1.27  &  III \\
01-Aug-02 07:42:21 &  992 - 3580 &    20 -    72 &   -6.01  &  DCIM \\
17-Aug-02 08:56:51 &  481 - 1300 &    10 -    90 &   -0.37  &  III \\
21-Aug-02 17:22:00 & 1060 - 3520 &    11 -    56 &   -2.02  &  DCIM \\
31-Aug-02 14:20:44 &  694 - 1865 &    13 -    70 &   -0.40  &  III RS \\
31-Aug-02 14:22:42 &  381 -  912 &    13 -   102 &   -0.33  &  III \\
27-Sep-02 13:02:07 & 1270 - 3610 &    16 -    57 &    0.46  &  DCIM \\
29-Sep-02 06:36:10 & 2080 - 3700 &    16 -    74 &   -0.25  &  DCIM \\
29-Sep-02 06:40:49 &  703 - 1240 &    17 -    58 &    0.01  &  DCIM \\
29-Sep-02 06:42:07 &  230 -  470 &    17 -    58 &   -0.52  &  III \\
22-Feb-03 09:28:16 &  685 - 2440 &    16 -    68 &    0.11  &  III \\
22-Feb-03 09:28:28 & 2050 - 3760 &    16 -    68 &   -1.36  &  III RS \\\hline
\end{tabular}
\caption{The final record of correlated events. A negative time delay indicates
that HXR comes before radio. The windows for correlation are given in center time,
radio frequency and HXR energy. The labels in the last column refer to (reversed) 
type III(RS), U bursts, and decimetric pulsations or patches (DCIM).}
\label{eventlist_tab}
\end {table}

\subsection{Radio and hard X-ray observations}
 
Phoenix-2 is a full-sun polarization-sensitive radio spectrometer located at Bleien (8$^o$6'44''E, 47$^o$20'26''N),
Switzerland. In its present configuration, it has a 7m dish and 200 narrowband frequency channels 
($\Delta \nu$ = 1, 3, or 10 MHz) covering the range from 0.1 to 4 GHz. Detection (after two IF stages) is 
logarithmic, and the nominal radiometric noise is 1 - 5\%.
The radio data were calibrated and cleaned from telecommunication artifacts using gliding background subtraction,
and bad channels were eliminated. Only radio flux density is considered for correlation with RHESSI data, but circular 
polarization was consulted for radio type identification. The radio spectrogram is integrated over a finite 
bandwidth in order to obtain a single time profile. 

The RHESSI satellite detects individual photons between 3 keV and 17 MeV in 4096 energy channels
with a time resolution of $2^{-20}$s ($\approx$ 1 $\mu$s). Although the full energy response matrix is
available for solar sources, we use here only the diagonal response because as our 
emphasis is on time structures in rather broad energy bands, and not on exact spectroscopy. RHESSI
is designed as a HXR imager, with two-dimensional imaging achieved by rotational modulation 
(\opencite{schnopper68}; \opencite{skinner95}). In this technique, the spatial information is encoded
in temporal modulation of the observed HXR flux when the source becomes shadowed by linear 
grids which are fixed on the rotating ($T_S$ $\sim$ 4s) spacecraft \cite{hurford02a}.
The RHESSI optics consists of 9 pairs of linear grids (`subcollimators') with angular pitches
$p_i = 2.61 \times 3^{i/2}$ arc seconds ($i$ = 1..9), and instantaneous modulation 
periods range from 5$\cdot$10$^{-4}$s to 2s, depending on subcollimator and source location.

\subsection{\label{instr_time_sect}Instrumental timing accuracy}

The Phoenix-2 clock is locked to GPS timing, accurate to one millisecond. The spectrometer uses UTC, derived from GPS time 
having no leap seconds. The timing of the $n$th frequency channel is shifted by $(n - 1)\times 0.5$ ms relative to the first 
channel, for which there is time stamp available on each data record. The time stamp in Phoenix-2 data has been measured by 
receiving the DCF77 time signal transmitted from Mainflingen near Frankfurt/Main (Germany) at 77.5 kHz. The absolute timing of 
DCF77 at emission is of the order of one millisecond. After correcting for propagation time, the Phoenix-2 time stamp was 
found delayed by 22$\pm$2 ms. This systematic radio delay is caused by signal propagation in the spectrometer and has been 
stable over time. It is corrected in the following where relevant, making the absolute Phoenix-2 timing accurate to within 
about 5 ms.

The RHESSI time stamp of each photon in UTC is set within approximately one $\mu$s of its arrival time. RHESSI 
timing is also synchronized with GPS and accurate within one millisecond (J. McTiernan, priv. comm.). The observations of 
more than a dozen non-solar gamma-ray bursts and soft X-ray repeaters have allowed to verify the RHESSI timing accuracy 
(K. Hurley, priv. comm.). A further systematic effect arises from the light travel time between RHESSI and Phoenix-2. 
The extreme cases arise when Phoenix-2 is at dusk/dawn and RHESSI is at noon, or vice versa. The maximum possible delay is 
thus  $\pm r_{\rm E}/c$ = $\pm$ 20 ms. It can go in both ways. The light travel time from RHESSI (altitude 600 km) 
to ground is only 2 ms, and always negative (radio delayed). Time differences due to effects in the solar corona will 
be discussed later.

\subsection{\label{decomp_vis_sect} Visibility-based HXR demodulation}

Since the RHESSI modulation interferes with temporal structures below the spin period, the RHESSI light curves 
must be demodulated prior to comparison with radio observations. Demodulation is an inverse problem, and
is here accomplished by the method of (Arzner 2002, 2004). In this `visibility-based' method, the solar HXR distribution during 
some 10 $T_S$ in a fixed energy band is assumed to be of the form $B(x,y,t) = \sum_k r_{t,k} B_k(x,y)$, 
and the binned time profile $r_{t,k}$ of the $k$-th source component is estimated by penalized maximum-likelihood 
\cite{yu94}. The goal quantity is the Bayesian probability ${\sf P}_{\rm tot} = {\sf L} \times {\sf P}_{\rm a}$, where ${\sf L}$ is
the posterior probability (likelihood) that the data are observed if the model was true, and ${\sf P}_{\rm a}$ is the 
a priori probability assigned to the model. The likelihood is calculated assuming Poisson statistics, and the a priori 
probability is chosen as
\begin{equation}
{\sf P}_{\rm a} = \exp \, \Big\{ - \frac{1}{2} \sum_{t,k} \alpha_k (r_{t+1,k}-r_{t,k})^2 \Big\} \, .
\label{P_a}
\end{equation}

The form of Eq. (\ref{P_a}) favors smooth time profiles without imposing a bias. The larger the value of $\alpha_k$, the smoother 
is $r_{t,k}$ as a function of $t$; as a rule-of-thumb, the autocorrelation time\footnote{defined by
$\tau^{-2}$ = $\int \hspace{-.5mm} S(\omega) \, \omega^2 \, d \omega \, \big/ \int \hspace{-.5mm} 
S(\omega) \, d \omega$ with $S(\omega)$ the power spectral density.} of the maximum-${\sf P}_{\rm tot}$ solution is
\begin{equation}
\tau_k \sim \max \, (\Delta t, \sqrt{\alpha_k} \langle c \rangle / \langle a_0 L\rangle) \;\;\;\; [s]
\label{tau_k}
\end{equation}
with $\Delta t$ being the time bin, $\langle c \rangle$ the average counts per second and subcollimator,
$a_0 \sim 0.25$ the mean subcollimator transmission \cite{hurford02a}, and 0 $\le L \le 1$ the detector 
lifetime \cite{schwartz02}. Angular brackets represent averages over time and subcollimators.
Different source components are thus identified by different intrinsic time 
scales, which allows an empirical separation of gradual and impulsive components. 

Although our Ansatz for $B(x,y,t)$ covers, in principle, arbitrarily 
complicated brightness distributions, we must restrict ourselves in practice to a few source
components ($N_k$=1,2,3) in order to avoid too ill-posed situations. As a consequence, 
we cannot account for arbitrarily fast moving sources. The method only works if
the source motion can be approximated by a (`movie') sequence of $N_k$ brightness distributions 
$B_k(x,y)$ within the subcollimator resolution. This requires that the source displacement 
during $1/N_k$-th of the integration time $T$ must not exceed 90$^o$ grid phase:
$v T / N_k < p_i/4$, where $v$ is the source velocity [asec/s], and
$p_i = 2.61 \times 3^{i/2}$ is the $i$-th angular pitch [asec]. Inserting $v$ = 0.13''/s 
= 100 km/s for (fast) footpoint motion, $N_k$=2, and $T$ = 42s (10 spin periods), we
find that subcollimators 
3 to 9 may be used. This represents a somewhat conservative constraint, because 0.13''/s is an upper
limit, and because the count rate rarely allows to resolve modulation in the
finest subcollimator \#1 (\#2 is not used due to high background).
If, however, there is no modulation observed in subcollimator \#1 then there
is also no conflict arising from footpoint motion.

Non-solar (non grid-modulated) background is accounted for
by allowing a constant offset at each subcollimator.
Since spatial and temporal observables of $B(x,y,t)$ are entangled, the ($\Delta t$-integrated) 
visibilities\footnote{Projections of $B_k(x,y)$ on the sine and cosine
components of the modulation pattern \protect\cite{hurford02a}.} must be estimated together with $r_{t,k}$.

Our numerical procedure to find the maximum-${\sf P}_{\rm tot}$ solution
is based on iteration. Starting from a flat time profile $r_{t,k}$ and zero visibilities, 
the code performs a sequence of partial Newton-Marquardt steps, with the Hessian
approximated by its diagonal, except for $r_{t,k}$, where the nearest-neighbour coupling (Eq. \ref{P_a}) is treated
by a full tridiagonal solver. Convergence is controlled by monitoring the goal function $\log {\sf P}_{\rm tot}$
and its contributions $\log {\sf L}$ and $\log {\sf P}_a$. The solution usually becomes stable after a 
few 100 iterations. Using the Portland Group Fortran 90 compiler with standard optimization on a medium-size 
work station, execution time for 500 iterations is about 1 second for typical time bins $\Delta t$ = 0.12 s, and
scales linearly with $1/ \Delta t$. 

While we make here no explicit use of the visibilities, they absorb, unavoidably,
observational information and thereby degrade the quality of $r_{t,k}$. An estimate on the
error of $r_{t,k}$ can be obtained by perturbing the solution $r_{t,k}$ until $\log {\sf L}$ deviates by 
more than $\frac{1}{2}$ from its unperturbed value (e.g., \opencite{eadie71}, \opencite{press98}).
The resulting error band is shaded gray in Figure \ref{closeup_fig} below;
it marks the range in which the probability of the observation, given the demodulation,
has dropped by a factor $e^{-1} = 0.36$. For consistency, the perturbations must have similar
time resolution as the unperturbed solution; this is enforced by keeping the perturbations 
constant in intervals $\tau_k$ (Eq. \ref{tau_k}).

\subsection{\label{accuracy_section}Timing accuracy of the demodulation}

A crucial issue for the data analysis is the timing reliability of the RHESSI demodulation. First, we discuss the reconstruction 
of the time profile at a resolution below the spin period. The error propagation into structure timing will be discussed later. It
should be stressed at this point that the error of the demodulation is usually dominated by systematic,
not stochastic sources. Exceptions arise at very low count rates, where the counting noise becomes
comparable to the uncertainty in the instantaneous RHESSI response due to the unknown source morphology. While 
the fast modulation is easily integrated out, the slow, `chirpy', modulation at glancing
RHESSI grid passages is much harder to estimate, and makes the demodulation problem ill-posed.

In order to assess the timing accuracy of the demodulation we pursue two strategies. The first one is based
on simulations and represents the intrinsic accuracy of the method. The second one
is based on true data and therefore includes realistic systematic errors due to violation of the model assumptions for
$B(x,y,t)$, use of an energy-averaged instrumental response, and imperfect background estimation.

\begin{figure}[ht]
\centerline{\epsfig{file=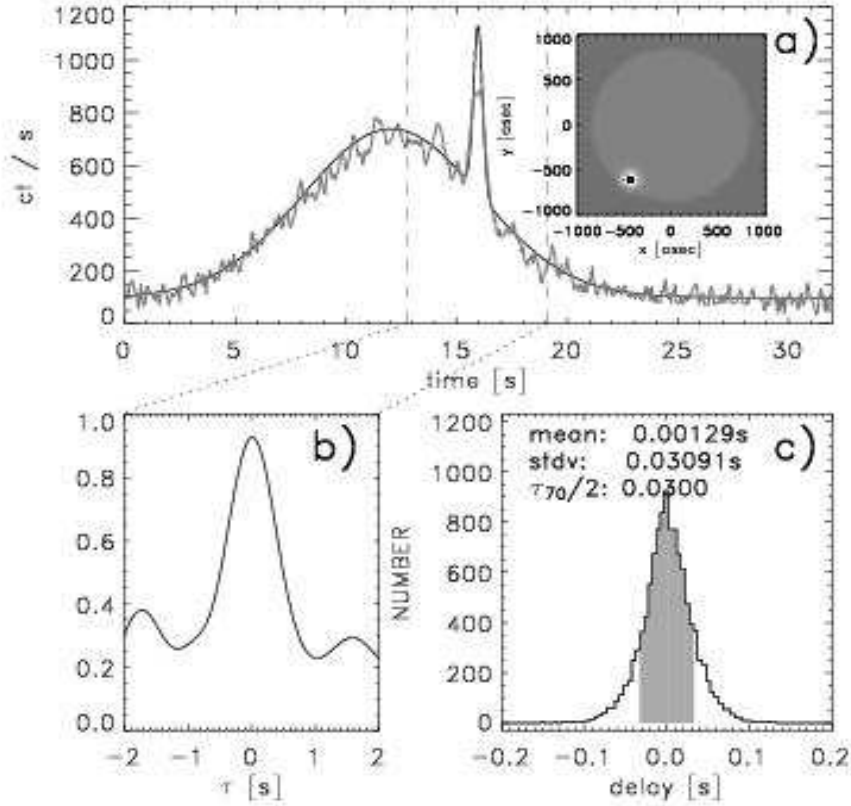,height=12cm,width=12.5cm}}
\caption{Simulation of the intrinsic timing accuracy of the demodulation. a) -- true time 
profile (black) and demodulation (gray) of a multicomponent Gaussian source (inlet). 
b) -- cross-correlation between true and demodulated time profiles. c) -- simulated 
delays of $10^4$ samples; the shaded range contains 70\% probability mass. See text.}
\label{sim_delay_fig}
\end{figure}

\subsubsection{\label{intrinsic_sect}Intrinsic accuracy of the demodulation method}

Let us start with the intrinsic timing accuracy. Figure \ref{sim_delay_fig} displays a test with $10^4$
simulated observations where both the model assumptions and the instrumental response are exact.
In each simulation, a simple source model is created (top, inlet), consisting of a constant background, 
active region (bright gray spot), and impulsive component (cross). Both the active region and the impulsive source are
modeled as Gaussians in space and time. The active region is placed at random on the solar disc, and the impulsive
source is placed at random within the active region. Using a simulated aspect solution and simulated data gaps 
(rate 1$s^{-1}$, mean duration 0.25s), the Poisson intensity is computed for each subcollimator, and a sequence of binned 
counts is generated. From these, the demodulation is calculated (Fig. \ref{sim_delay_fig}a gray line) 
and compared to the true spatially integrated profile (Fig. \ref{sim_delay_fig}a black line). The role of
the active region is to provide a time-dependent background with a generally non-vanishing slope, which
may, potentially, bias the timing. Such a slope-induced bias is thus
included in the simulation. 
The simulated average count rate of Fig. \ref{sim_delay_fig} is 350 counts/s/subcollimator, 
which is representative for an M class flare. The reconstructed curve has
a time resolution 
$\Delta t$ = 0.062s and smoothing parameters $\alpha_1 = 10^{-1}$ and $\alpha_2 = 3\cdot10^{-4}$. 
The cross-correlation with the original, in the dashed interval of panel a),
is presented in panel b). 
Panel c) shows the distribution of $10^4$ simulated delays between the original and the 
reconstructed curves, as given by the peak cross-correlation coefficient. 
The delays have a standard deviation $\sigma_{\rm intr}$ = 0.031s, 
where the subscript refers to the intrinsic accuracy of the demodulation method. Alternatively, we may consider the 
shaded region in Fig. \ref{sim_delay_fig}c, which contains 70\% probability mass and has width $\tau_{\rm 70}$ = 0.060s.
For Gaussian statistics, one expects $\tau_{\rm 70}/2 = \sigma$; this is well fulfilled for the simulation of Fig.
\ref{sim_delay_fig}, but deviations may occur for real data (Sect. \ref{systematic_sect}). In this case, $\tau_{\rm 70}$
is the more significant quantity.

class flares with larger count rates allow finer time bins, smaller smoothness parameters $\alpha_k$, 
and better intrinsic timing accuracy $\sigma_{\rm intr}$. From simulations similar to Fig. \ref{sim_delay_fig}
we deduce the estimate
\begin{equation}
\sigma_{\rm intr} \simeq \frac{0.3 s}{\sqrt{\langle c \rangle}} \;\;\;\;\; \mbox{if} \;\;\; \langle c \rangle \ga 100 \, {\rm s}^{-1}
\label{sigma_intr}
\end{equation}
where $\langle c \rangle$ is the average count rate in ct/s/subcollimator in the whole time interval ($\sim$ 10 $T_S$) 
under consideration. (It is not the peak intensity alone which matters because modulation is to be identified
from the whole time interval.) The limit $\langle c \rangle \ga 100 \, s^{-1}$ in Eq. (\ref{sigma_intr})
is understood in the sense that above this value, $\sigma_{\rm intr}$ becomes weakly dependent on the model characteristics
such as the total signal-to-background count ratio or the instantaneous signal-to-background intensity; below
100 ct/s/subcollimator, the achievable accuracy depends on the model characteristics and is typically
lower ($\sigma_{\rm intr}$ $\sim$ 0.1 s). We may thus interpret a count rate of 100 ct/s/subcollimator as a minimum
requirement for reliable use of the visibility-based demodulation method. It should be recalled that Eq. (\ref{sigma_intr}) 
refers to the timing accuracy as measured in the numerical experiment of Fig. \ref{sim_delay_fig},
and does not include any systematic (instrumental) errors.

A similar error as $\sigma_{\rm intr}$ must be expected from a cross-correlation with a radio signal assumed to be without noise. 
Although the error introduced by demodulation exceeds the instrumental timing errors (Sect. \ref{instr_time_sect}), 
it is symmetric and thus cancels out on the average. 

\begin{figure}[ht]
\centerline{\epsfig{file=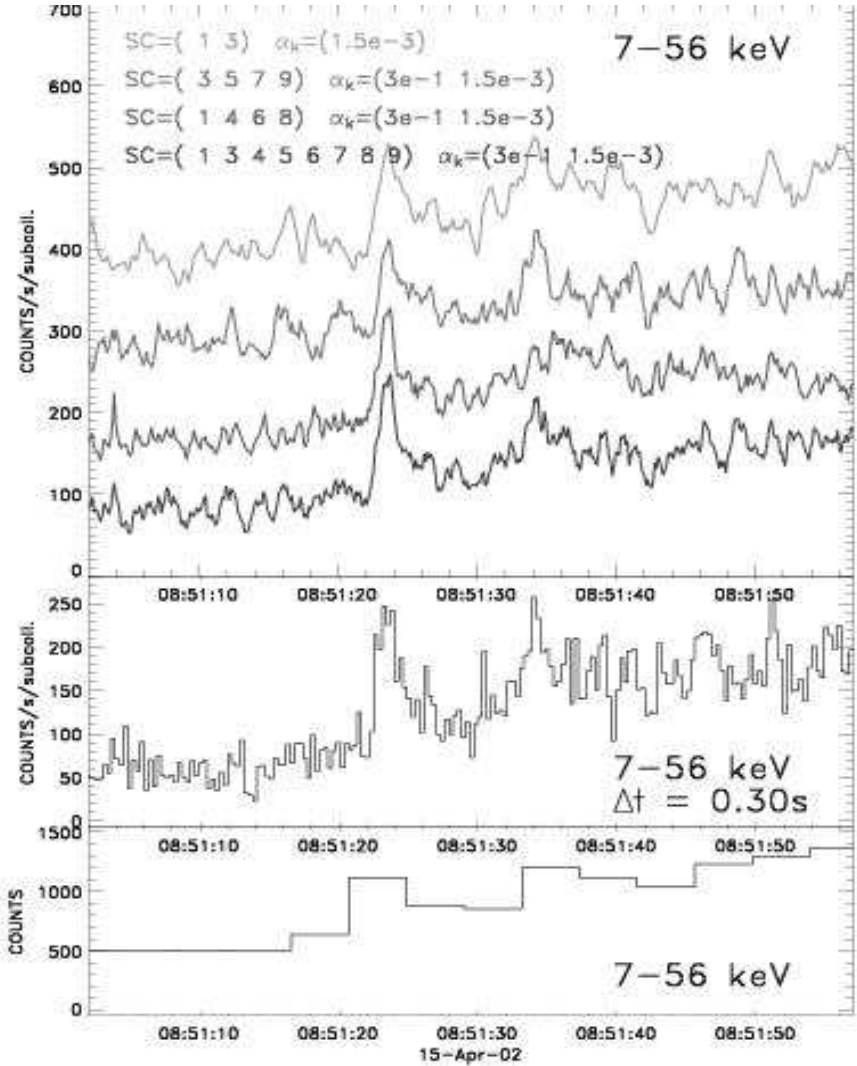,height=14.5cm,width=12cm}}
\caption{Top panel: different solutions of the demodulation problem, using different subcollimator (SC) 
sets and smoothness parameters $\alpha_k$. Subsequent graphs are offset by 100 ct/s/subcollimator 
for better clarity, with the bottom graph having zero offset. Middle panel: an alternative demodulation method 
(private comm. G. Hurford). Bottom panel: raw data, in time bins of one spin period.}
\label{solutions_fig}
\end{figure}

\subsubsection{\label{systematic_sect}Systematic errors}

We turn now to the second strategy, involving real data.
An estimate for the robustness of demodulated features may be gained by varying the subsets of subcollimators
and smoothness parameters $\alpha_k$, and by comparison with other demodulation methods.
Figure \ref{solutions_fig} (top panel) shows the demodulation results from different sets of
subcollimators and $\alpha_k$ parameters using data of the flare of April 15, 2002, 08:51:30.
Different curves are offset by 100 ct/s for better clarity, with the bottom curve having zero offset.
The non-solar background is forced here to zero in order to facilitate the comparison with an alternative 
demodulation method (middle panel) which does not include non-solar background.
The top curve of Fig. \ref{solutions_fig} is obtained from the finest two subcollimators under the 
assumption of a single source component. The next two curves represent odd-numbered (3,5,7,9) and even-numbered 
(1,4,6,8) subcollimators, assuming two source components in both cases (collimator \#2 is replaced by \#1 due
to increased background). Both subcollimator sets cover the full 
range of angular resolutions and grid orientations. The bottom curve involves all subcollimators except \#2. 
Generally, more subcollimators contain more information and therefore should give better estimates. However,
pieces of information from different subcollimators may sometimes be contradictory. This is, for instance,
the case at 08:51:34 for even and odd subcollimators (top panel, middle curves). It may also happen that
only one single subcollimator sees a short intense peak. The global demodulation then
barely responds, since the likelihood cost for all other subcollimators would be too high.
Erratic contributions from cosmic rays are (hopefully) suppressed in this way.
The middle panel of Figure \ref{solutions_fig} shows the result of an alternative demodulation method proposed by 
G. Hurford (priv. comm.). This method removes sinusoidal contributions at the instantaneous modulation frequency, 
inferred from the source centroid, and corrects for time-dependent grid transmission and detector lifetime. The algorithm 
acts locally in time in intervals of duration 0.3s, which were chosen to be commensurate with 
the time resolution (Eq. \ref{tau_k}) of the visibility-based demodulation. The middle panel 
is to be compared to the bottom curve of the top panel, as both represent the true incoming counts per second and 
subcollimator. Finally, the bottom panel presents the raw counts in bins of one spin period ($T_S = 4.1428$s). 
This may serve as a very coarse but trustworthy benchmark. 

\begin{table}[ht]
\begin{tabular}{l|lllll}
 	    &	${\tt O}_1$  	& ${\tt O}_2$ 		& ${\tt E}_1$ 		& ${\tt E}_2$ 	& ${\tt A}_2$ \\\hline     
${\tt O}_1$ &	             	&       0.014 (0.010)  & 	0.396 (0.420)	& 	0.408 	(0.427) & 0.056 (0.026) \\
${\tt O}_2$ & {\it -0.0007}	&	 	     	& 	0.365 (0.348)	& 	0.392 	(0.404) & 0.034 (0.031) \\
${\tt E}_1$ & {\it  0.067}	& {\it	0.045}   	& 	 		& 	0.012   (0.0005) & 0.016 (0.012) \\
${\tt E}_2$ & {\it  0.043}	& {\it	0.011}   	& {\it  0.0008} 	&		& 0.033 (0.022) \\
${\tt A}_2$ & {\it  0.014}	& {\it	0.012}  	& {\it -0.004} 		& {\it -0.0035} &      
\end{tabular}
\caption{Above diagonal: standard deviation and $\tau_{\rm 70}/2$ (in brackets) of the delay between
different demodulations. Below diagonal: corresponding mean values. All values are in seconds. The labels
${\tt O}$ and ${\tt E}$ refer to even and odd-numbered subcollimators, and 
the subscripts denote the number of assumed source components. ${\tt A}$ refers to all subcollimators 
except \#2. Each entry is computed over a sample of 54 HXR peaks.}
\label{estim_err_sys_tab}
\end {table}

The top panel of Figure \ref{solutions_fig} suggests a method to determine the total (intrinsic + systematic) timing accuracy of the 
visibility-based demodulation. By selecting manifest peaks like at 08:51:23 UTC, and correlating the demodulations of different
subsets of subcollimators and varying $\alpha$ parameters, an estimate on the effective timing accuracy can be obtained. 
This procedure gives an upper bound; the timing error with respect to the relatively more accurate radio 
signal should be smaller by a factor $\sim 2^{-1/2}$. We have carried out the above programme for a total
of 54 identified HXR peaks during the events listed in Tab. \ref{eventlist_tab}, employing 5 different combinations of subcollimator 
sets and $\alpha$ parameters. The result is summarized in Table \ref{estim_err_sys_tab}. 
The label ${\tt O}_1$ denotes the odd-numbered subcollimators (3,5,7,9) with a single source component; ${\tt O}_2$
the odd-numbered subcollimators with two source components; ${\tt E}_1$ the even-numbered subcollimators (1,4,6,8) with one
source component, ${\tt E}_2$ similarly with two components. Finally, ${\tt A}_2$ involves all subcollimators (except \#2) and
two source components; it is ${\tt A}_2$ which is normally used for the correlation with the radio data.
The smoothness parameters $\alpha_k$ were chosen such that single and
double sources had a similar time resolution, which also corresponds to the one used in the HXR-radio correlation study.
The entries of Tab. \ref{estim_err_sys_tab} above the diagonal are the standard deviation and $\tau_{\rm 70}/2$ (in brackets) in seconds,
where $\tau_{\rm 70}$ is defined empirically as in Fig. \ref{sim_delay_fig}c. Below the diagonal is the mean delay in seconds. There are two major
features which become apparent from Tab. \ref{estim_err_sys_tab}: first, the introduction of additional source components
(${\tt O}_1$ $\to$ ${\tt O}_2$ and ${\tt E}_1$ $\to$ ${\tt E}_2$) has only a minor effect on the demodulation solution,
and the resulting delays are within the intrinsic accuracy (Sect. \ref{intrinsic_sect}). Secondly, the use of disjoint
observational data sets (${\tt O}_k \leftrightarrow {\tt E}_k$) leads to much larger discrepancies in the order of 0.4s.
Not surprisingly, ${\tt A}_2$ agrees better with each of its subsets ${\tt O}_2$ and
${\tt E}_2$ than these do among each other. The mean values (lower triangle) are of the order of the corresponding standard 
deviations divided by $\sqrt{54}$, so that we may consider them as consistent with zero. This argument is, however,
qualitative only because the statistics is not strictly Gaussian ($\tau_{\rm 70}/2 \not= \sigma$).
Whereas the standard deviation between 
even and odd subcollimators ($\sim$ 0.4s) certainly over-estimates the absolute error of the full estimator ${\tt A}_2$, 
the deviation between ${\tt A}_2$ and the others ($\sim$0.04s) under-estimates it because of overlapping data sets.
The overall standard deviation of all pairs of $({\tt O}_1,{\tt O}_2,{\tt E_1},{\tt E}_2,{\tt A}_2)$ is found to be
0.17s. The authors argue that the latter value, sited between the intrinsic error $\sigma_{\rm intr}$ and the
worst-case estimate derived from odd versus even subcollimators, represents a reasonable and conservative estimate on 
the absolute timing accuracy of the RHESSI demodulation.

\section{\label{compare_sect}Radio - HXR Comparison}

\subsection{Procedure}

\begin{figure}[ht]
\centerline{\epsfig{file=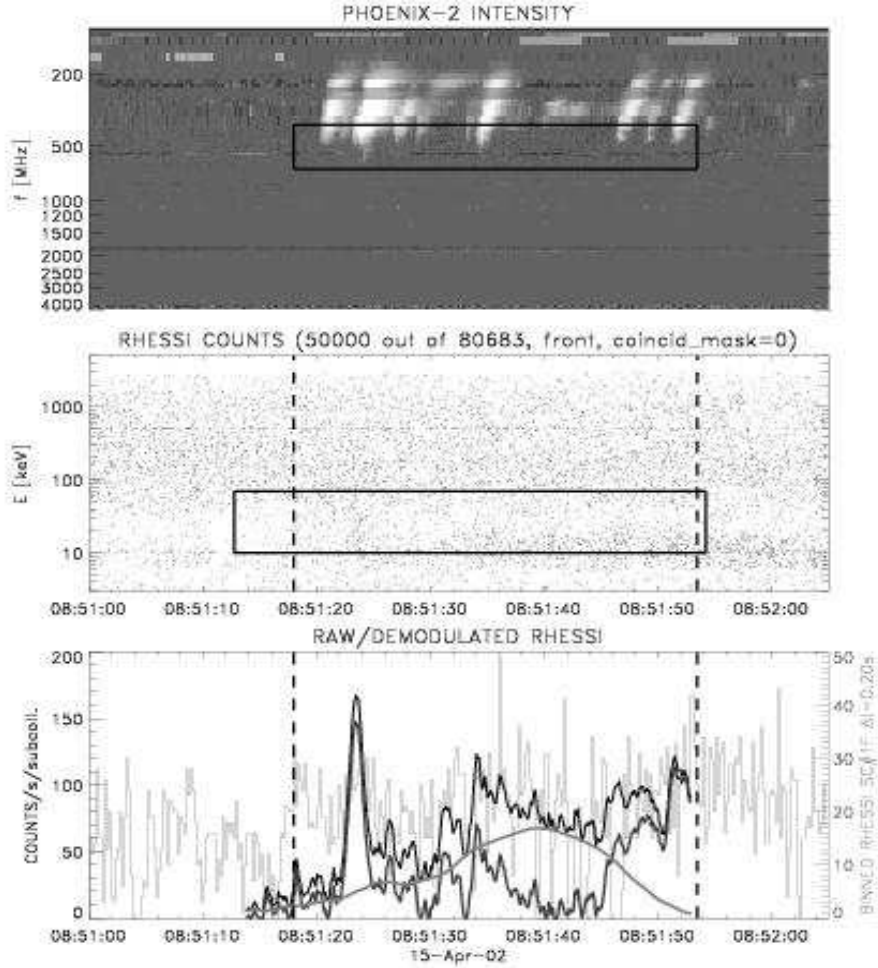,width=12.5cm,height=13cm}}
\caption{Top: radio spectrogram, and time-frequency box selected for cross-correlation. Middle: RHESSI raw counts 
(all subcollimators; only a random subset of 50.000 is shown) and time-energy window (solid line) selected for demodulation.
Bottom: subcollimator \#1, binned in 0.2s bins (histogram style)
and demodulated time profiles (light gray: slow component; dark gray: fast component; black: slow+fast components).}
\label{overview_fig}
\end{figure}

Once the radio and HXR data are calibrated and demodulated, respectively, they are confronted with each other and searched for 
common fine structures. We illustrate our procedure in Figure \ref{overview_fig}, considering as example the GOES C class 
flare of April 15, 2002, 08:51 UTC. In the top and middle panels of Figure \ref{overview_fig}, the calibrated Phoenix-2 
spectrogram is displayed against the RHESSI raw counts. Grayscale represents histogram-equalized logarithmic radio flux. Dots represent 
a subset of all observed RHESSI counts, selected at random to show the raw distribution. In a first step, we select a 
HXR energy band and time range containing the bulk of flare-associated counts (middle panel, solid line; Table \ref{eventlist_tab}). 
The average count rate in this energy band is then used to choose time bins (see \opencite{arzner04} for 
details; here: $\Delta t$ = 122 ms), and a binned event list is generated. This, together with the 
aspect solution (\opencite{fivian02}, \opencite{hurford02b}) and grid transmission \cite{hurford02a}, forms the input to 
the demodulation code. By varying the subset of subcollimators, the reliability of temporal 
structures is tested, and different choices of smoothness constraints $\alpha_k$ are explored 
(Section \ref{accuracy_section}, \ref{solutions_fig}). 
The demodulation $r_{t,k}$ adjudged optimum is shown in the bottom panel of Figure \ref{overview_fig}.
It has two source components with time scales $\tau_1 \sim $ 8.7s (light gray) and $\tau_2 \sim $ 0.5s (dark gray),
which add up to the total solar HXR flux (black). All subcollimators 
were used, except for \#2, suffering enhanced background \cite{smith02}. For comparison, 
the binned ($\Delta t$ = 0.2s) raw counts of the finest subcollimator \#1 are also shown (Fig. \ref{solutions_fig} 
bottom, histogram style). Since 0.2s exceeds the modulation frequency of subcollimator \#1 (except at glancing 
source passages of the RHESSI grids), this is expected to yield a noisy proxy for the true time profiles. 
Data gaps \cite{smith02}, where the count rate drops to zero, are not corrected in the binned raw counts, 
but are incorporated -- via detector lifetime -- in the demodulation. With some 100 ct/s/subcollimator, 
the event is among the weaker ones considered in this study. Contrary to Fig. \ref{solutions_fig}, the
non-solar is not kept zero but fitted as 35\%.

\begin{figure}[ht]
\centerline{\epsfig{file=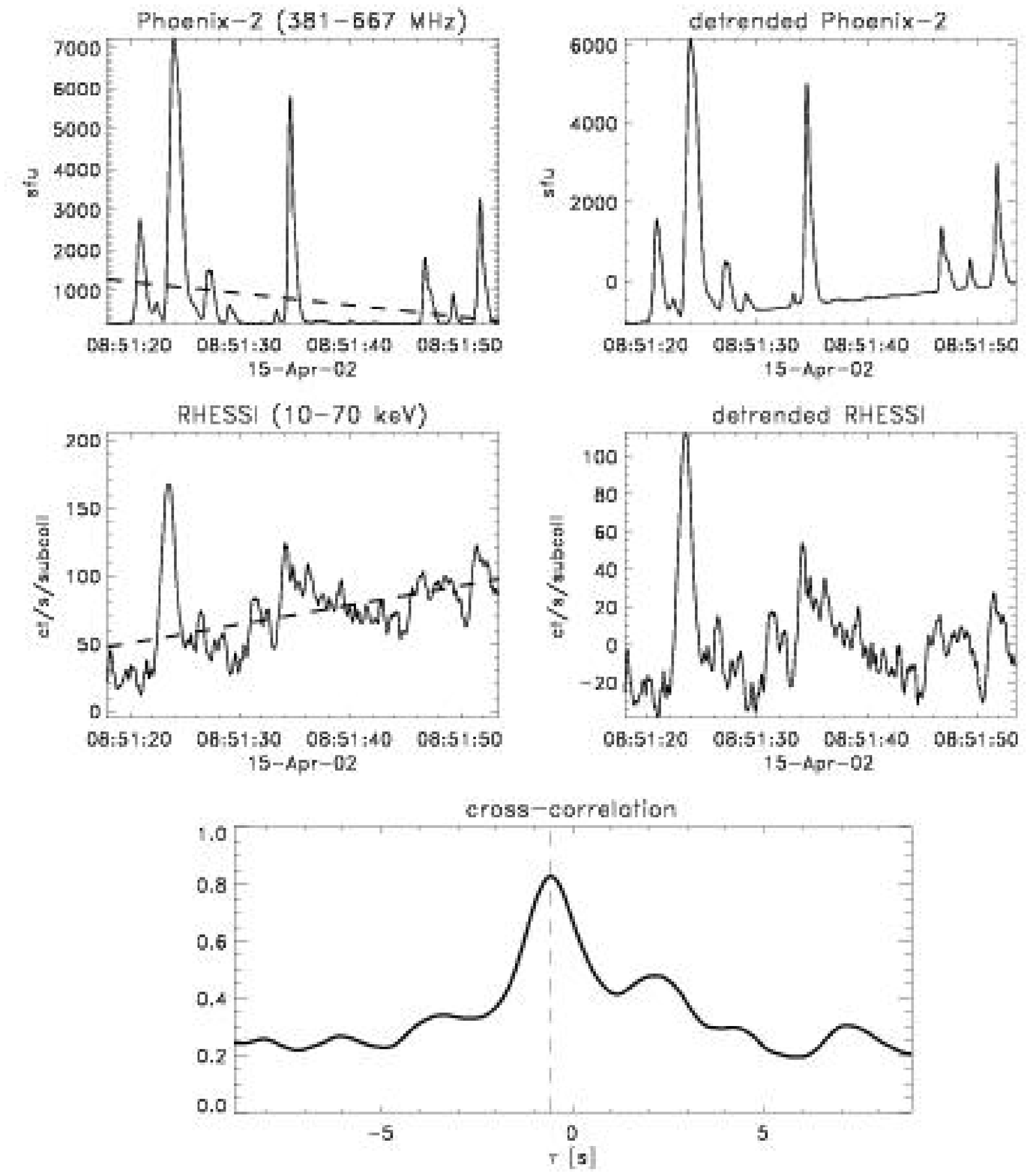,width=10.5cm,height=12cm}}
\caption{Close-up of Fig. \protect\ref{overview_fig}: time profiles of Phoenix-2 (top line) 
and demodulated RHESSI (middle line). The RHESSI signal represents the total flux summed over both source components. 
A linear trend (left column, dashed)
is subtracted to obtain de-trended light curves (right column), the cross-correlation of which
is shown in the bottom panel. Negative $\tau$ indicates that HXR comes first. The error of
the demodulation, obtained by numerical perturbation of the solution,
is shaded gray (left column, middle row).}
\label{closeup_fig}
\end{figure}

In a next step, we select a time-frequency box (Fig. \ref{overview_fig} top panel, solid line; Table \ref{eventlist_tab}) 
containing the radio emission to be correlated with the HXR. In doing so we attempt to capture the onset of 
type IIIs such as to minimize delays due to the radio drift. The RHESSI light curve is the sum of all 
source components if several are used. The radio spectrogram inside the selected 
time-frequency box is then frequency-integrated to obtain the radio light curve shown in
Fig. \ref{closeup_fig} (top left). From this, a linear trend is subtracted (Fig. \ref{closeup_fig} top right),
and the result is cross-correlated with a similarly de-trended RHESSI light curve (Fig. \ref{closeup_fig} middle 
row). The HXR de-trending amounts to a removal of slow variations, which typically arise from
low-energy (thermal) contributions.
The cross-correlation is shown in the bottom panel of Fig. \ref{closeup_fig}. The time delay is
defined from the peak of the cross-correlation, and indicated by dashed line. The gray shaded region in
the left middle panel is an error estimate of the demodulation (see Sect. \ref{decomp_vis_sect}).
By comparing the lightcurves in the right column of Fig. \ref{closeup_fig} we conclude that
a chance coincidence of the two major peaks is unlikely, but a one-to-one correspondence of all structures
is certainly contestable. With a leap of imagination one may draw
associations between individual RHESSI and Phoenix-2 peaks, and investigate their relative
timing. Averaging would then yield a mean delay which is compatible with
the maximum of the cross-correlation of the full time series.

\begin{figure}
\centerline{\epsfig{file=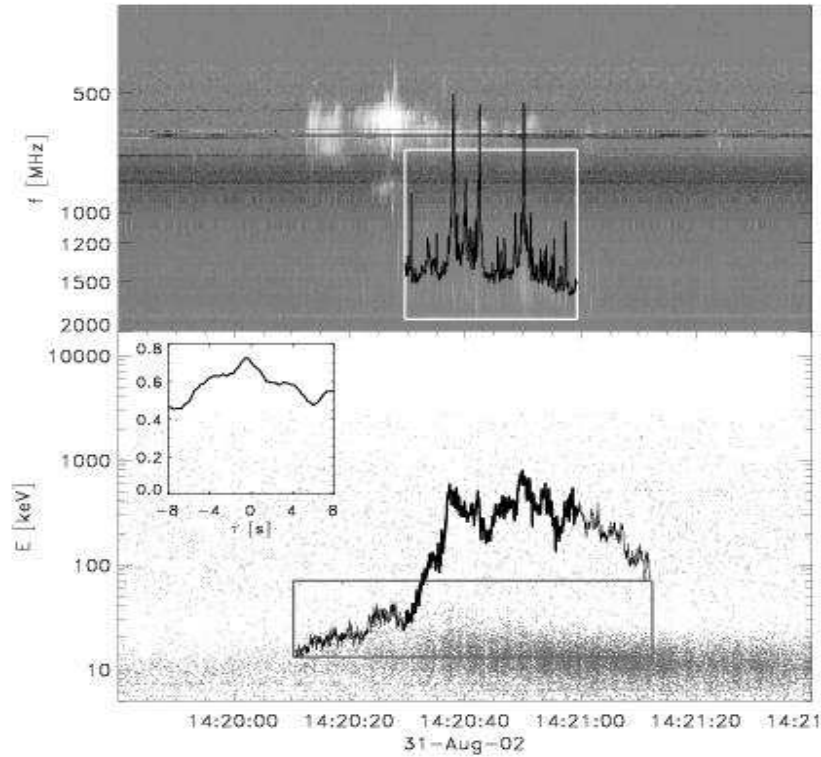,width=11cm,height=10.5cm}}
\caption{The event of August 31, 2002, 14:20 UTC.}
\label{20020831142000}
\end{figure}

\begin{figure}
\centerline{\epsfig{file=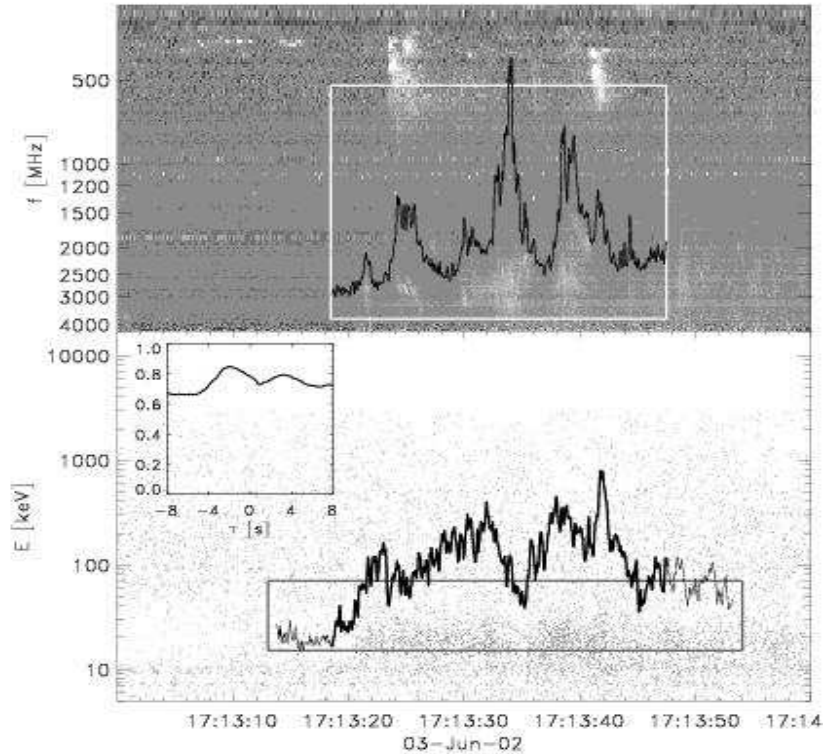,width=11cm,height=10.5cm}}
\caption{The event of June 3, 2002, 17:13 UTC.}
\label{20020603171300}
\end{figure}

\begin{figure}
\centerline{\epsfig{file=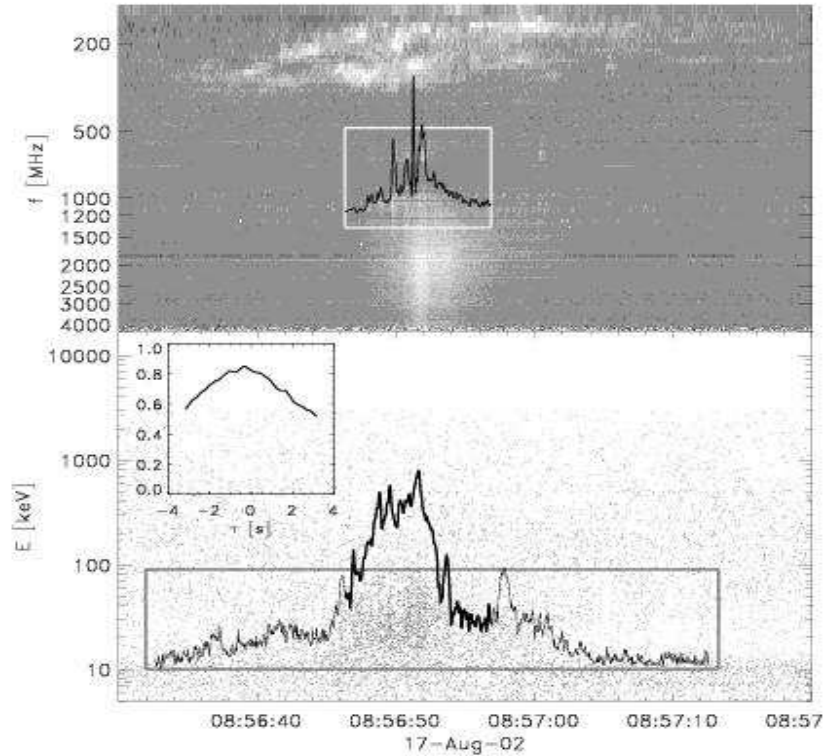,width=11cm,height=10.5cm}}
\caption{The event of August 17, 2002, 08:56 UTC.}
\label{20020817085630}
\end{figure}

\begin{figure}
\centerline{\epsfig{file=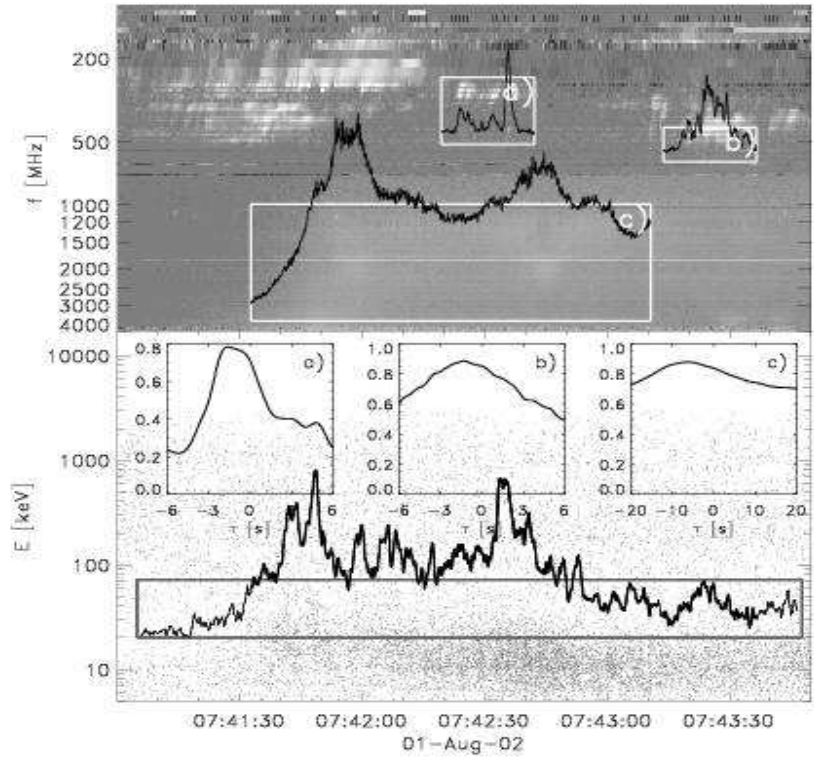,width=11cm,height=10.5cm}}
\caption{The event of August 1, 2002, 07:41 UTC.}
\label{20020801074100}
\end{figure}

\begin{figure}
\centerline{\epsfig{file=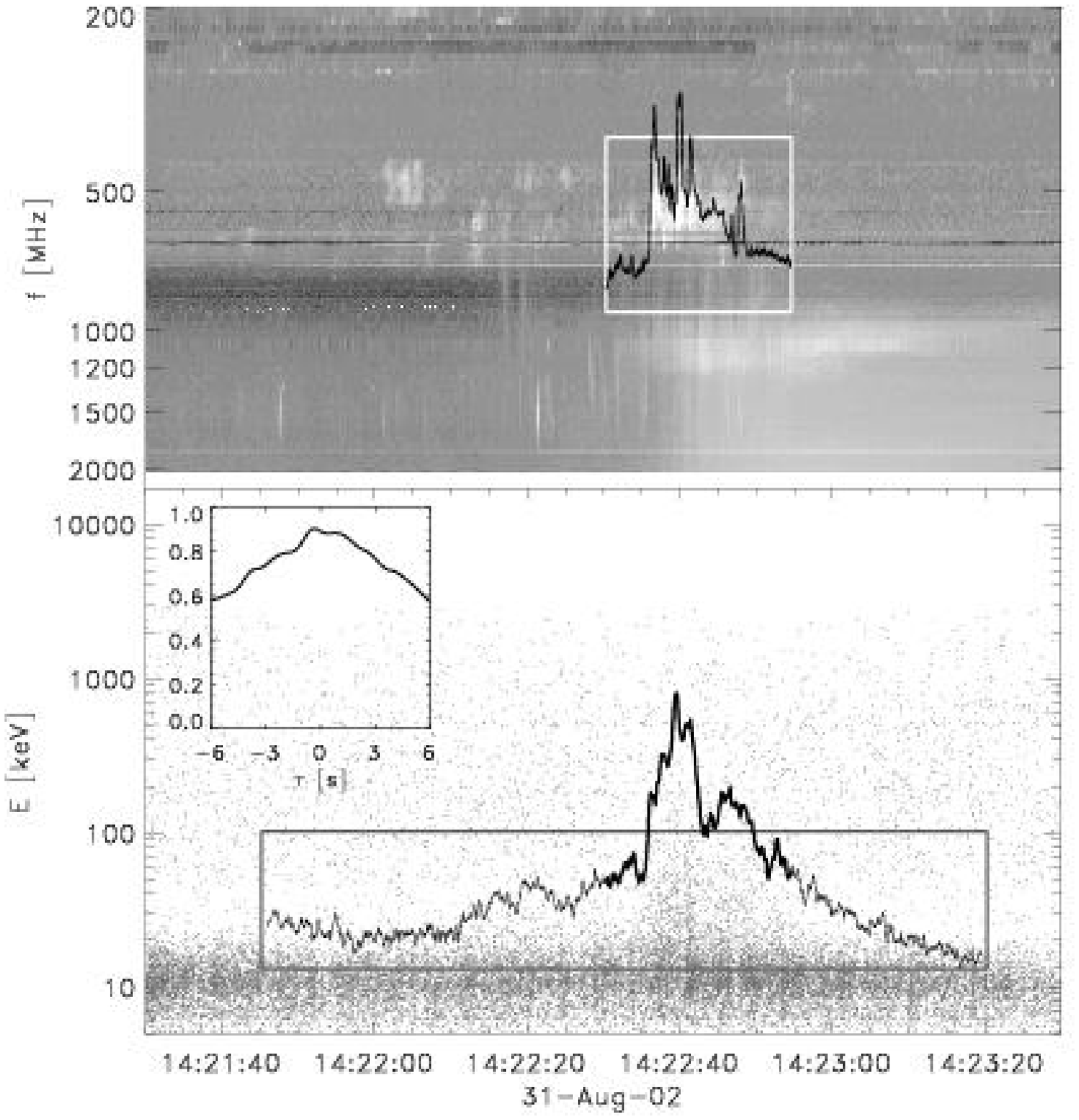,width=11cm,height=10.5cm}}
\caption{The event of August 31, 2002, 14:22 UTC.}
\label{20020831142200}
\end{figure}

\begin{figure}
\centerline{\epsfig{file=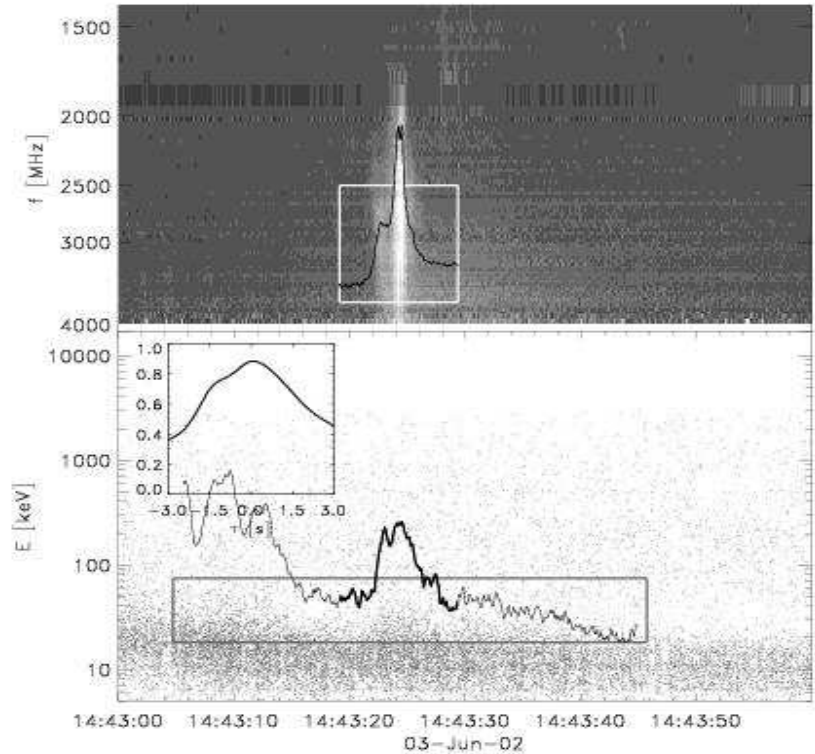,width=11cm,height=10.5cm}}
\caption{The event of June 3 2002, 14:43 UTC.}
\label{20020603144300}
\end{figure}

\begin{figure}
\centerline{\epsfig{file=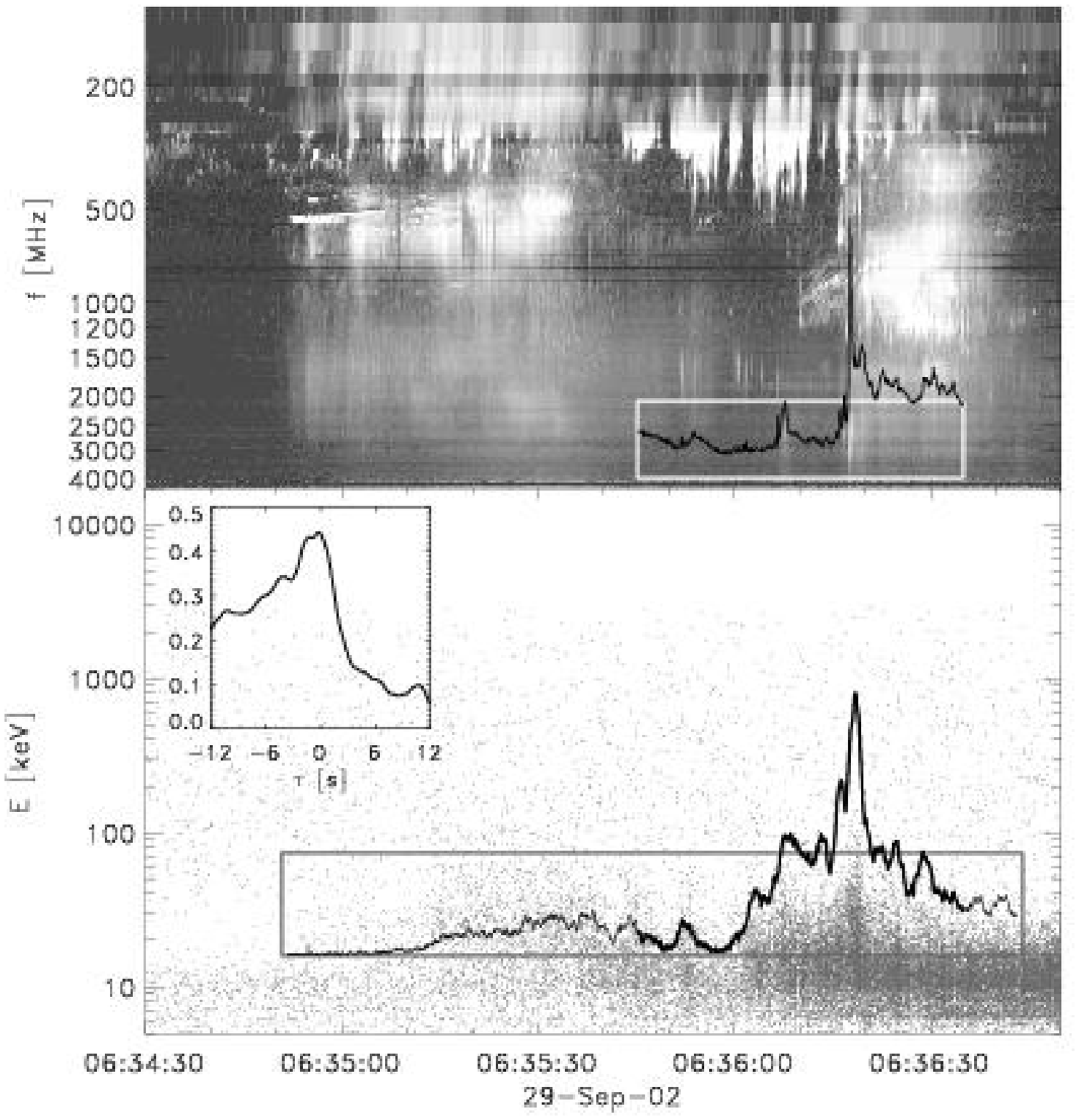,width=11cm,height=10.5cm}}
\caption{The event of September 29, 2002 06:36 UTC.}
\label{20020929063610}
\end{figure}

\subsection{Survey}

\begin{figure}
\centerline{\epsfig{file=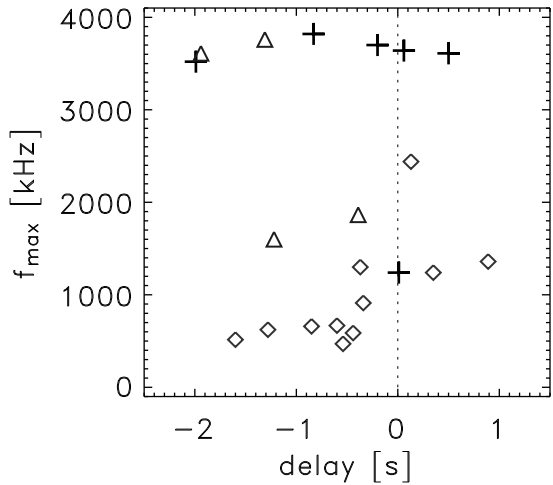,width=8cm,height=7cm}}
\caption{Time delay of the radio emission (against HXR) versus maximum radio frequency. Different symbols
refer to normal drifting type III bursts ($\diamond$), reversed drifting type III bursts ($\triangle$),
and decimetric pulsations or patches ($+$).}
\label{scatter_fig}
\end{figure}

All events listed in Table \ref{eventlist_tab} have been processed in the way described above. The radio burst types have 
been classified according to \inlinecite{isliker94}. They include type III bursts and DCIM, an abbreviation used in Solar and 
Geophysical Data for events in the decimetric range that do not fit the metric classification, such as pulsations, patches 
and narrowband spikes. The delays were determined by cross-correlation as described above. A total of 24 bursts has been 
analyzed. Their average delay is $-$0.85$\pm$0.28 s, thus the HXRs are generally leading. The time delay is plotted in
Fig. \ref{scatter_fig} versus the maximum frequency used in the analysis (Fig. \ref{overview_fig} top, rectangle).

The 17 groups of correlating type III bursts are delayed by $-$0.69$\pm 0.19$ s on average. The distribution is broad, 
having a standard deviation of 0.79 s. Only 3 out of 15 have a positive delay. The 8 groups of type III bursts with 
maximum frequency above 1 GHz are delayed less on average ($-$0.45s) than the groups below 1 GHz ($-$0.96 s). Although the standard 
deviations are so large that these trends are not statistically significant, they exist independently and are significant 
within groups (e.g. Fig. \ref{20020603171300}). In particular, it can be seen from Fig. \ref{scatter_fig} that the type 
III delays (diamonds) below 1 GHz are consistently negative, while at higher frequencies positive delays occur as well.
Surprisingly, the 4 groups of reversed slope type III bursts (triangles) are also delayed (average $-$1.2 s). It should be 
pointed out that not all of the investigated type III events have clear HXR correlations on the level of individual peaks.
Judged by eye, we estimate that only some 20\% of the XHR-associated type III bursts correlate peakwise; 
this estimate is limited by the small statistics and by the sensitivity of the HXR demodulation.

The average delay of the 7 DCIM bursts (Fig. \ref{scatter_fig}, crosses; the outlier at $-$6.01s is not shown)
is $-$1.25$\pm$0.86 s, but the distribution is broad, having a standard deviation 
of 2.27 s. Thus 2 out of 7 DCIM have positive delay, and the mean delay of DCIM is statistically not different from zero. 
The order of magnitude of the delays is similar to the one reported by \inlinecite{Aschwanden92} for decimetric 
narrowband spikes, but we do not find a correlation of the delay with HXR peak flux.

In reality, the individual events are too different to allow simple statistics. In the following we present some selected 
examples proceeding from cases of type III-only events to pure DCIM emissions.

\begin{itemize}

\item 31-Aug-02, 14:20:20 (Fig. \ref{20020831142000}): In radio waves, this event consists of just three groups of type III bursts. 
A sequence of normal drifting type IIIs around 600 MHz is extremely intense, reaching 7054 sfu at 623 MHz. A short group of type III at 
800 MHz includes some reversed slope bursts. However, what correlates with HXR in their period of maximum flux is a sequence of short 
reverse drifting type III at 1500 MHz starting at 14:20:25. The HXR images resolve a single source only. The demodulation with two source 
components (not shown), however, reveals an interesting delay, with the rise of the slower component being roughly the time integral of faster one 
(but energies are the same, contrary to Neupert effect). 

\item 3-Jun-02 17:13:30 (Fig. \ref{20020603171300}): The radio emission consists of just reversed slope type III bursts. They occur in 
two groups of different frequencies. The correlation with X-rays is best if both radio emissions are included. However, the two bursts 
at 500 MHz are delayed by $-$1.92s, whereas the group around 3 GHz is delayed by only $-$0.35 s. 

\item 17-Aug-02 08:56:50 (Fig. \ref{20020817085630}): The radio event starts with a type II-like emission at meter waves, includes type III at 
decimeter waves, and a decimetric patch from 1.2 to beyond 4 GHz with opposite circular polarization. The type III bursts seem to be 
recorded as fundamental emission around 660 and as harmonic emission at 1100 MHz. Fundamental/harmonic pairs exceeding 1 GHz are reported here 
for the first time. The fundamental band is more delayed (by about $-$0.1 s) than the harmonic band. The decimetric patch is 2s later than HXR. 

\item 1-Aug-02, 07:41 (Fig. \ref{20020801074100}):  The rich radio event starts with metric type III bursts and includes oppositely polarized 
decimetric patches during the main HXR phase. There is rough agreement between the broad HXR and DCIM structures, with HXR first 
by 6 s. Contrary to the low-frequency group of type III bursts, two later groups of decimetric type III bursts correlate with HXR. 
This is at times when the DCIM emission shows no temporal structures.  

\item 31-Aug-02, 14:22:30 (Fig. \ref{20020831142200}): The event has a preflare phase of weak HXR, during which several groups of 
decimetric bursts occur. The correlation does not become obvious until the main phase, when it is best with a group of type III bursts 
around 570 MHz. At about the same time, however, a decimetric patch becomes very intense, but continues much longer than the HXR emission. 

\item 3-Jun-02 14:43:25 (Fig. \ref{20020603144300}): A decimetric patch occurred after a C3.7 flare. Such `afterglows' are often observed
after large flares. The patch was preceded by slowly drifting reversed slope type III bursts at the same frequencies. Both seem to 
correlate well HXR in the 20-80 keV range.

\item 29-Sep-02, 06:35:40 (Fig. \ref{20020929063610}). Two similar events were recorded at 06:35:00 and 06:36:20 UTC in radio waves, 
each with intense normal-drifting metric type III bursts, a highly polarized type II-like  burst at relatively high frequency (starting 
beyond 500 MHz), and DCIM (up to 3.5 GHz). The first event is accompanied with few HXR. In the second event, the HXRs seem to correlate 
better with the DCIM emission than with the various groups of type III bursts.

\end{itemize}

\section{Discussion}

The comparison of the light curves of coherent radio emission and flare HXRs is puzzling. In a majority of the events there seems to be 
no correlation at all. Of the 24 correlating segments 17 involve type III bursts. The rest are broadband pulsations or patches. 
These DCIM bursts show a larger scatter in the cross-correlation delay than the type III bursts. A rather small fraction
($\sim$ 20\%) of the selected HXR-associated type III bursts show peakwise correlation with HXR. The delays were found to be rather robust
against change of the HXR energy band, and in particular against a moderate increase of the lower bound to exclude thermal contributions. 
This is mostly due to the trend removal applied before the correlation (Fig. \ref{closeup_fig}), which suppresses the gradual evolution.
Increasing the lower energy bound tends to slightly {\it in}crease the delay, as would be
expected from a residual (positive, thermal) slope that shifts the HXR peak maximum towards later times.

The lack of correlation or a shift in the timing of the two emissions may originate from different acceleration 
sites of the emitting particles \cite{benz05}. Let us assume in the following that the correlating radio 
and HXR emissions are caused by the same electron population and discuss possible origins of delays. 
The {\it instrumental errors} are $<$5 ms for timing accuracy and $<$20 ms for instrument positions (Section 1). 
The demodulation causes a scatter of about 170 ms (Section 2). The latter two errors cancel on average, but dominate 
in a small sample.

There are many sources of {\it delays at the origin} of the radiation discussed in the following.

1) Type III sources at 300 MHz are located at an average height of 2.2$\cdot 10^5$km \cite{Paesold01}. The majority of the HXR sources 
are at footpoints of field lines in the chromosphere. This simple geometry predicts a delay of the HXRs by 0.74 s in 
the center of the disk, but vanishes at the limb. It contradicts the observed sign of the delay and absence of a 
center-to-limb effect in the delay. 

2) As electrons propagate from the acceleration site to the origin of radiation, the emission is delayed by 
\begin{equation}
\Omega t \ = \ {l_x\over v_x} - {l_r\over v_r} 
\end{equation}
where $l_x$ and $l_r$ are the distances from the acceleration site to the HXR source and radio source, respectively. 
The electrons causing the emission travel with velocities $v_x$ and $v_r$. For 25 keV electrons, producing typically 
15 keV photons, $v_x=9.3\cdot 10^9$cm/s. Assuming $v_x\approx v_r$, the maximum time delay is -2.4 s if acceleration 
is at zero altitude and coincides with the HXR source. The particle propagation delay balances the source location delay 
at disk center if acceleration occurs at an altitude of $7.3\cdot 10^9$cm. Acceleration below this altitude lets the particle 
propagation delay dominate, causing a negative net value.

3) The radio emission is further delayed by its group velocity being smaller than the speed of light. Using the dispersion 
relation for electromagnetic waves, $\omega^2 = c^2 k^2 + \omega_p^2$, the group delay relative to propagation with 
the speed of light amounts in first order of $(\omega_p/\omega)^2$ to
\begin{equation}
\tau \ = \ {{2\pi e^2}\over {c m_e \omega^2}}\int n_e ds\ =\ {H_n\over 2c\cos \theta}\left({\omega^0_p\over\omega}\right)^2
\end{equation}
\cite{benz02}, where $\theta$ is the radiation propagation angle relative to vertical (assumed constant) and $\omega^0_p$ is 
the plasma frequency at the radio source. For a density scale hight $H_n=10^{10}$cm, appropriate to a temperature 
of 2 MK, the group delay (emitted at the plasma frequency $\omega^0_{p}$) is 0.17/cos$\theta$ for the fundamental, 
and 4 times less for the harmonic. The effect may thus be relevant for fundamental emission and for harmonic sources near the limb. 

4) Radio scattering in the solar corona is another source of systematic time delays. Scattering leads to angular
source broadening, and radiation from the observed halo travels a longer distance than radiation from the 
observed core. Assuming free propagation between the scattering screen and the observer,
the resulting time delay is $\tau_f = (1-D_1/D) \langle  (\Delta \phi)^2 \rangle / (2c)$, 
where $D_1$ is the distance from the source to the scattering screen, $D$ is the distance from the source to the observer, 
and $\Delta \phi$ is the observed source radius \cite{benz02}. Assuming $D_1 = 10^7$m and using the observed minimum 
source radius for $\Delta \phi$ \cite{bastian94} gives $\tau_f$ = 8.8 ms at 1.4 GHz and 35 ms at 330 MHz. 
A second type of scattering delay is due to signal detours {\it within} the scattering screen.
For a source embedded in a screen of inhomogeneous plasma, and within the geometric-optics approximation, the radio pulse is delayed 
(and broadened) by $\tau_i \sim \eta (L/c\bar{n})^2$ \cite{arzner99}, where $\bar{n}$ is the average refractive index, $L$ is the 
distance from the source to the screen surface (from where the radiation freely propagates to the distant observer), and $\eta$ 
is the continuous-time angular diffusion coefficient of the geometric optics rays. For smooth electron density fluctuations 
$\delta n$ with Gaussian two-point function of correlation length $l$ one finds 
$\eta = 4^{-1} \sqrt{\pi}c l^{-1} (1-\bar{n}^2)^2 \bar{n}^{-3} (\delta n / \bar{n})^2$.
Assuming $l \sim 10^5$m, $L \sim D_1 \sim 10^7$m, $\delta n / \bar{n} \sim 0.1$, and fundamental emission at
$f = 1.05 \times f_p$ = 500 MHz, the scattering delay is $\tau_i \sim 0.5$s. Both $\tau_f$ and $\tau_i$ are thus potentially 
relevant contributors to the observed radio delay.

\section{Conclusions}

The study has shown that RHESSI data can be demodulated sufficiently accurately to allow timing between HXR emission and 
radio bursts. Although an uncertainty is introduced into the delay of peak cross-correlation, it does not affect the average. 
Structures in coherent radio emissions of solar flares, in particular of type III and decimetric pulsations and patches 
(DCIM), occasionally correlate with HXRs, but are generally delayed. The scatter of the delay, measured by cross-correlating 
the two emissions, is large. Its sign is negative (radio delayed) in a large majority of type III bursts, and its average is 
statistically significantly negative by a few 100 ms. The large scatter in delays may indicate that several effects work in 
opposite ways. The delays by different source location, particle propagation, radio group velocity and scattering, all in 
the few 100 ms range, make this plausible. 

Comparing the observed delays with the mechanisms discussed in the previous section, we conclude:

\begin{itemize}

\item The absence of a center-to-limb effect suggests that the source location delay does not play an important role.
 
\item Similarly, the group velocity delay cannot be fully responsible for the delay of the radio emission unless it is 
concentrated to the immediate environment of the radio source and thus independent of the viewing angle. Under this 
condition it may explain the delay of fundamental emission observed in type III bursts (Fig. \ref{20020817085630}).

\item The fact that delayed reverse drift type III bursts have been observed several times suggests that the particle 
propagation delay does not seem to be a major cause for delayed radio emission. 

\item Scattering near the source thus remains as the prime delay mechanism. It does not produce much of a center-to-limb 
effect and is larger for fundamental emission as observed. Delay by scattering is also consistent with the generally 
observed decrease at higher frequencies (Fig. \ref{20020603171300}).

\end{itemize} 

The timing of DCIM structures relative to HXR shows a broader distribution in delays than for type III bursts. Although the 
average is negative, the delays including 4 positive cases out of 7 DCIM events. In other words, the correlation is less 
tight in DCIM than type III bursts (Fig. \ref{20020801074100}). These findings do not confirm a general delay of pulsations 
and patches as previously reported for narrowband spikes, but do not corroborate a close correlation of DCIM emission and particle 
acceleration neither.

Timing structures of radio bursts and HXR emission yields information on acceleration and emissions. As the different 
source locations and propagation times are major contributors to the delay, accurate timing is potentially important to further unravel 
the relation of radio and HXR emissions in flares. The limiting factor of this study was not the HXR demodulation, but the 
lack of radio imaging. This may become available in the future through the Frequency Agile Solar Radiotelescope (FASR).

\begin{acknowledgements}
The authors thank D. Quinn for help with extensive data handling, G. Hurford, A. Csillaghy, 
P. Saint-Hilaire and the RHESSI team for helpful discussions, and Ch. Monstein for supervision of Phoenix-2 
observations. The Phoenix-2 and RHESSI work at ETH Zurich are supported, in part, by the Swiss National 
Science Foundation (grant nr. 20-67995.02) and ETH Z\"urich (grant nr. TH-W1/99-2).
\end{acknowledgements}

\end{article}

\end{document}